\documentclass[11pt,fleqn]{article}

\usepackage{amsmath,amssymb,amsthm,enumerate}

\newcommand{\p}{\partial}
\newcommand{\const}{\mathop{\rm const}\nolimits}

\newcommand{\sign}{\mathop{\rm sign}\nolimits}

\newcommand{\todo}[1][\null]{\ensuremath{\clubsuit}}

\newcommand{\noprint}[1]{}

{\theoremstyle{definition} 
\newtheorem{example}{Example}

\begin{document}

\par\noindent {\LARGE\bf
More common errors in finding exact solutions\\ of nonlinear differential equations. I
\par}

{\vspace{4mm}\par\noindent 
Roman O. POPOVYCH$^{\dag\ddag}$ and Olena O.~VANEEVA$^{\ddag}$
\par\vspace{2mm}\par}

{\vspace{2mm}\par\noindent\it 
${}^\dag$Fakult\"at f\"ur Mathematik, Universit\"at Wien, Nordbergstra{\ss}e 15, A-1090 Wien, Austria
\\[1mm]
${}^\ddag$Institute of Mathematics of NAS of Ukraine, 3 Tereshchenkivska Str., Kyiv-4, Ukraine
\\[1mm]
\rm E-mail: \it  rop@imath.kiev.ua, vaneeva@imath.kiev.ua
\par}

\noprint{
{\vspace{9mm}\par\noindent\hspace*{8mm}\parbox{140mm}{\small
Recently Kudryashov [{\it Commun.\ Nonlinear Sci.\ Numer.\ Simulat.}, 2009, {\bf 14}, 3507--3529] 
listed seven common errors in finding exact solutions of differential equations. 
We indicate two more errors concerning point equivalence and linearizability of differential equations. 
The first error is discussed and exemplified with many papers on generalized KdV and mKdV equations.  
}\par\vspace{7mm}}
}

{\vspace{5mm}\par\noindent\hspace*{8mm}\parbox{140mm}{\small
In the recent paper by Kudryashov [{\it Commun.\ Nonlinear Sci.\ Numer.\ Simulat.}, 2009, {\bf 14}, 3507--3529] 
seven common errors in finding exact solutions of nonlinear differential equations were listed and discussed in detail. 
We indicate two more common errors concerning the similarity (equivalence with respect to point transformations) 
and linearizability of differential equations and then discuss the first of them. 
Classes of generalized KdV and mKdV equations with variable coefficients are used in order to clarify our conclusions.  
We investigate admissible point transformations in classes of generalized KdV equations, 
obtain the necessary and sufficient conditions of similarity of such equations to the standard KdV and mKdV equations
and carried out the exhaustive group classification of a class of variable-coefficient KdV equations.
Then a number of recent papers on such equations are commented using the above results. 
It is shown that exact solutions were constructed in these papers only for equations 
which are reduced by point transformations to the standard KdV and mKdV equations. 
Therefore, exact solutions of such equations can be obtained 
from known solutions of the standard KdV and mKdV equations in an easier way than by direct solving. 
The same statement is true for other equations which are equivalent to well-known equations with respect to 
point transformations.
}\par\vspace{3mm}}

\noprint{
}

\section{Introduction}

Finding exact (resp.\ closed-form, resp.\ explicit) solutions plays a significant role 
in investigation of physically important partial differential equations, especially, nonlinear ones. 
There exist a number of famous named solutions in different fields of physics. 
First such solutions were constructed and intensively studied for models describing flows of fluids 
(e.g., the Navier--Stokes and Euler equations and the boundary layer equation)  
since it is difficult to neglect nonlinear effects arising in fluid dynamics. 
These were the Poiseuille flow, the Burgers vortex,
Jeffery--Hamel solution for the flow in a wedge shaped region,   
the von Karman solution for flow over a single rotating disc etc.\ 
(see, e.g., the extensive reviews by Berker and by Pukhnachev \cite{Berker1963,Pukhnachev2006}).
Simultaneously with penetrating the nonlinear paradigm into other fields of science, 
the construction of exact solutions became more and more topical and involved nonlinear diffusion and wave equations and systems of such equations, 
nonlinear Schr\"odinger equations, general relativity equations etc. 
Some equations have only a few known exact solutions (e.g, the Ablowitz--Zeppetella traveling wave solutions of the Fisher equation), 
for other equations wide families of exact solutions parameterized with a number of arbitrary constants and functions have been obtained 
(integrable and linearizable equations and equations possessing large Lie symmetry groups). 
Last decades the scientific activity in this field permanently grows. 

Recently \cite{Kudryashov2009} Kudryashov has made an excellent review and 
listed \emph{seven common errors} in finding exact solutions of nonlinear differential equations, 
which appeared in the modern mathematical and physical literature. 
The above errors relate to correctness, generality and representation of such solutions and methods of their constructions. 
In spite of its carefulness, the list by Kudryashov does not exhaust all existing common errors in this subject. 
We indicate two more common errors, continuing numbering of~\cite{Kudryashov2009} for convenience of further references. 
These errors usually arise on the very initial step of consideration, 
namely, under the choice of differential equations for finding exact solutions and related investigations.

\begin{description}
\item[Eighth error.]
Exact solutions are often constructed with no relation to equivalence of differential equations 
with respect to point (resp.\ contact, resp.\ potential etc.) transformations.
\item[Ninth error.]
Exact solutions of linearizable differential equations are sometimes found with no relation to the linearizability property.
\end{description}

Mathematical investigations are always carried out up to equivalence relations defined in sets of similar objects. 
Equivalent objects differ only in properties which are inessential for certain consideration and, therefore, 
can be identified in certain sense. 
This rule is extended to mathematical models, arising in applications. 

A usual equivalence used for differential equations is that generated by locally nondegenerate point transformations 
and called \emph{similarity}~\cite{Ovsiannikov1982}. 
Similar systems of differential equations have similar local properties and similar related local objects 
(local solutions, symmetries, local conservation laws etc.).  

The purpose of this letter is to discuss the eighth error. 
In fact this error can be considered as a whole family of different \emph{inaccuracies concerning similarity of differential equations}. 
We distinguish, e.g., the following kinds of inaccuracies.
\begin{enumerate}
\item
Any differential equation from the class under consideration is similar to the same classical differential equation 
for which wide multiparametric families of exact solutions were constructed earlier and 
possibilities for finding new solutions inequivalent to known ones look as problematic. 
\item
Only differential equations from a subclass of the class under consideration are similar to classical ones 
and exact solutions are obtained only for such equations. 
Sometimes constraints singled out the subclass of equations with found particular solutions from the whole class 
are explicitly presented, sometimes they are implicitly implied or even not indicated in any way. 
\item
The class of differential equations under consideration can be mapped by a family of point transformations 
parameterized with arbitrary elements of the class to another class of simpler structure 
(resp.\ with less number of arbitrary elements, resp.\ with simpler transformational properties etc.) 
but this possibility is missed. 
\item
Similarity of differential equations is used for finding exact solutions but 
the usage is incomplete (resp.\ improper, resp.\ incorrect). 
\end{enumerate}    

There are rather a lot of published papers in which exact solutions were constructed for equations 
similar to well-known and/or simpler equations without using the similarity. 
Choosing examples for commenting, we restricted ourselves with quite recent papers published in certain journals 
and devoted to considerations of equations related to the KdV and mKdV equations. 
Our choice is justified by the fact that both the equations are the most known and well-investigated nonlinear equations 
of mathematical physics. 
They are connected via the Miura transformation.
Since these equations are integrable and have nice symmetry properties, 
wide families of their exact solutions had been found using different powerful methods 
(the inverse scattering method, Darboux and B\"acklund transformations, the Hirota bilinear method etc.)
and collected in a number of papers, textbooks and handbooks 
(see, e.g., \cite{Ablowitz&Segur1981,Matveev&Salle1991,Miura1968,Rogers&Shadwick1982,Sakhnovich1993} and references therein). 
At the same time, in the commented papers only very particular results on similar equations were obtained. 

The illustration of the above list of inaccuracies with generalized KdV and mKdV equations 
is additionally justified by their own importance in different fields of physics (see, e.g., references in commented papers). 
Moreover, the existence of similarity transformations between equations which are related to the KdV and mKdV equations 
is a well-known fact from the late 1960's.  
A~simple point transformation connecting the Gardner and mKdV equations was already presented in~\cite{Miura1968}.
A~more nontrivial point transformation between the standard and cylindrical KdV equations was found in~\cite{Lugovtsov&Lugovtsov1969} 
(see also~\cite{Hirota1979}). 
Then the similarity of certain generalizations of KdV and mKdV equations with coefficients depending on~$t$ or $(t,x)$ 
to these equations was investigated in detail (see, e.g., \cite{Grimshaw1979,Gurses&Karasu1994,Hirota1979,Hlavaty1988} and references therein).
Later the similarity arguments were permanently used under reviewing and refereeing papers similar to commented ones, 
cf.~\cite{MathRev1999}. 
Lie and generalized symmetries of such equations were also investigated 
(see, e.g., \cite{Gungor&Lahno&Zhdanov2004,Sergyeyev1999,Winternitz&Gazeau1992}).

Our paper is organized as follows: 
In the next section we construct 
a hierarchy of normalized classes of third order $(1+1)$-dimensional evolution equations, 
which is related to the examples commented.  
This gives the complete description of admissible point transformations within such classes. 
For two wide subclasses of the variable-coefficient KdV and mKdV equations, 
jointly covering equations from almost all the examples,
we derive the necessary and sufficient condition of similarity of such equations 
to the standard KdV and mKdV equations, respectively.
Exhaustive group analysis of a normalized class of variable-coefficient KdV equations 
is carried out in Section~\ref{SectionOnGroupAnalysisOfClassOfvcKdVeqs}.
Section~\ref{SectionOnExamplesOnSimilarityOfEqsAndTheirSolutions} is partitioned into two subsections. 
In subsection~\ref{SectionOnExamplesOnCompleteSimilarity} we collect examples 
when the whole considered classes of equations are similar to the KdV or mKdV equations, 
exemplifying the first of the inaccuracies listed above. 
Subsection~\ref{SectionOnParticularSimilarity} includes examples on particular similarity, 
which illustrate the second and third inaccuracies.
Different errors under the construction of similarity transformations among generalized KdV and mKdV equations 
are classified and illustrated by examples in Section~\ref{SectionOnInaccuraciesConcerningSimilarityTransformations}.
In the conclusion other possible applications of similarity transformations are discussed.

\section{Admissible point transformations in classes of generalized\\ KdV equations}
\label{SectionONAdmissiblePointTransInClassesOfGenKdVEqs}

Following~\cite{Popovych&Kunzinger&Eshraghi2009}, we start from the general class of third order evolution equations 
and construct a hierarchy of nested normalized subclasses of this class, which consist of different generalizations of the KdV and mKdV equations.
In this way we describe the entire sets of admissible point transformations of these subclasses.
At first we briefly define necessary notions, considering only the case of single differential equations in one unknown function for simplicity. 
Rigorous definitions for the general case of systems of differential equations are given, e.g., in~\cite{Popovych&Kunzinger&Eshraghi2009}.  

Let~$\mathcal L_\theta$ be a differential equation $L(y,u_{(p)},\theta(y,u_{(p)}))=0$ for the~unknown function $u$
of $n$~independent variables $y=(y_1,\ldots,y_n).$
Here $u_{(p)}$ denotes the set of all the derivatives of~$u$ with respect to $y$
of order not greater than~$p$, including $u$ as the derivative of order zero,
and $L$ is a fixed function depending on~$y,$ $u_{(p)}$ and~$\theta$.
The tuple $\theta$ of $k$ arbitrary (parametric) functions $\theta^1(y,u_{(p)})$, \ldots, $\theta^k(y,u_{(p)})$
runs through the set~$\mathcal S$ of solutions of the auxiliary system $S(y,u_{(p)},\theta_{(q)}(y,u_{(p)}))=0$
of differential equations with respect to $\theta$.
In this system $y$ and $u_{(p)}$ play the role of independent variables
and $\theta_{(q)}$ stands for the set of all the partial derivatives of $\theta$ of order not greater than $q$
with respect to the variables $y$ and $u_{(p)}$.
Usually the set $\mathcal S$ is additionally constrained by the condition
$\Sigma(y,u_{(p)},\theta_{(q)}(y,u_{(p)}))\ne0$ with another differential function~$\Sigma$.
In what follows we call the functions $\theta$ arbitrary elements.
Also, by~$\mathcal L|_{\mathcal S}$ we denote \emph{the class of equations~$\mathcal{L}_\theta$ with the arbitrary elements $\theta$ running through $\mathcal S$}.
The set~$\mathrm{T}(\mathcal L|_{\mathcal S})$ of triples $(\theta,\tilde\theta,\varphi)$, where 
$\theta,\tilde\theta\in\mathcal S$ and $\varphi$ is a point transformation from $\mathcal L_\theta$ to $\mathcal L_{\tilde\theta}$ 
if such a transformation exists, is called the {\em set of admissible transformations in the class~$\mathcal L|_{\mathcal S}$}.
(This is a formalization of the notion of allowed~\cite{Winternitz&Gazeau1992} or form-preserving~\cite{Kingston&Sophocleous1998} transformations.)
Roughly speaking, the (usual) \emph{equivalence group} $G^{\sim}$ of the class~$\mathcal L|_{\mathcal S}$ 
consists of the point transformations in the space of variables and arbitrary elements, 
which are projectable on the variable space and preserve the whole class~$\mathcal L|_{\mathcal S}$. 
The class~$\mathcal L|_{\mathcal S}$ is \emph{normalized} if any
admissible transformation in this class is induced by a transformation from~$G^{\sim}$.

Consider the general class $\mathcal E^3$ of third order evolution equations. They have the form 
\[u_t=H(t,x,u,u_x,u_{xx},u_{xxx}),\] where $H_{u_{xxx}}\ne0$.
To find the set of admissible transformations of the class~$\mathcal E^3$ and its complete equivalence group 
including both discrete and continuous equivalence transformations, we apply the direct method. 
The calculations are simplified with taking into account the well-known fact that 
the expression for~$t$ in any point (and, even, contact) transformation connecting two $(1+1)$-dimensional evolution equations depends only on~$t$
\cite{Kingston&Sophocleous1998,Magadeev1993}. 
Thus, a point transformation maps an equation from~$\mathcal E^3$ to an equation from the same class if and only if it has the form 
\begin{equation}\label{EqContactTransOfGenEvolEqs}
\tilde t=T(t), \quad 
\tilde x=X(t,x,u), \quad 
\tilde u=U(t,x,u).
\end{equation}
The functions~$T$, $X$ and $U$ have to satisfy the nondegeneracy assumption $T_t\Delta\ne0$, where $\Delta=X_xU_u-X_uU_x$.
The equivalence group~$G^{\sim}_0$ of~$\mathcal E^3$ consists of the transformations 
\begin{equation}\label{EqPointTransOfGenEvolEqs}
\tilde t=T(t),\quad \tilde x=X(t,x,u),\quad \tilde u=U(t,x,u),\quad 
\tilde H=\frac\Delta{T_tD_xX}H+\frac{U_tD_xX-X_tD_xU}{T_tD_xX},
\end{equation}
where $T$, $X$ and $U$ run through the corresponding sets of smooth functions satisfying the above nondegeneracy assumption 
and $D_x$ denotes the operator of total differentiation with respect to the variable~$x$, 
$D_x=\p_x+u_x\p_u+u_{tx}\p_{u_t}+u_{xx}\p_{u_x}+\cdots$. 
Therefore, the class~$\mathcal E^3$ is normalized. 

The subclass~$\mathcal E^3_{0.1}$ singled out from~$\mathcal E^3$ by the constraint $H_{u_{xxx}u_{xxx}}=0$ 
has the same equivalence group~$G^{\sim}_0$ and, therefore, is normalized. 
The same claim is true for the subclass~$\mathcal E^3_{0.2}$ of~$\mathcal E^3_{0.1}$, 
associated with the additional constraint $H_{u_{xxx}u_{xx}}=0$.

Setting one more constraint $H_{u_{xxx}u_x}=0$ leads to the normalized subclass~$\mathcal E^3_{1.1}$ of~$\mathcal E^3_{0.2}$
whose equivalence group~$G^{\sim}_1$ is a proper subgroup of~$G^{\sim}_0$ 
consisting of the transformations~\eqref{EqPointTransOfGenEvolEqs} with $X_u=0$. 
The normalized subclass~$\mathcal E^3_{1.2}$ singled out from~$\mathcal E^3_{1.1}$ by the constraint $H_{u_{xxx}u}=0$ 
has the same equivalence group~$G^{\sim}_1$.

The equivalence group~$G^{\sim}_2$ of the normalized subclass~$\mathcal E^3_2$ 
nested in~$\mathcal E^3_{1.2}$ and associated with the additional constraint $H_{u_{xxx}x}=0$ is properly contained in~$G^{\sim}_1$. 
Its elements additionally satisfy the condition~$X_{xx}=0$.

Another possibility is to impose the constraint $H_{u_{xx}}=0$ within the subclass~$\mathcal E^3_{0.1}$. 
The corresponding subclass~$\mathcal E^3_3$ is normalized and possesses the equivalence group~$G^{\sim}_3$ 
formed by the transformations~\eqref{EqPointTransOfGenEvolEqs} for which $X_u=0$, $U_{uu}=0$ and $2U_{ux}X_x=U_uX_{xx}$.

Then the subclass $\mathcal E^3_4=\mathcal E^3_2\cap\mathcal E^3_3$ is normalized 
and its equivalence group is~$G^{\sim}_4=G^{\sim}_2\cap G^{\sim}_3$. 
Integrating the total set of the constraints imposed on the arbitrary element~$H$ 
and the conditions obtained for elements of~$G^{\sim}_4$ gives that 
the equations from the class~$\mathcal E^3_4$ have the form 
\[
u_t+g(t)u_{xxx}=F(t,x,u,u_x),
\]
where $g\ne0$. 
The equivalence group~$G^{\sim}_4$ consists of the transformations (we present only components corresponding to the equation variables)
\begin{equation}\label{EqGenKdVEquivGroup4}
\tilde t=\alpha(t),\quad
\tilde x=\beta(t)x+\gamma(t),\quad
\tilde u=\theta(t)u+\Phi(t,x),\quad 
\end{equation}
where $\alpha$, $\beta$, $\gamma$, $\theta$ and $\Phi$ are arbitrary smooth functions of their arguments, $\alpha\beta\theta\ne0$. 

Further we consider two more special subclasses of~$\mathcal E^3_4$ which are directly related to 
examples commented in Sections~\ref{SectionOnExamplesOnSimilarityOfEqsAndTheirSolutions} 
and~\ref{SectionOnInaccuraciesConcerningSimilarityTransformations}.   

The first subclass is formed by the variable-coefficient KdV equations
\begin{equation}\label{EqvcKdV}
u_t+f(t)uu_x+g(t)u_{xxx}+h(t)u+(p(t)+q(t)x)u_x+k(t)x+l(t)=0,
\end{equation}
where all the parameters are arbitrary smooth functions of~$t$, $fg\ne0$. 
This subclass is normalized. 
Its equivalence group (in terms of the arbitrary element~$H$) is singled out from~$G^{\sim}_4$ by the condition $\Phi_{xx}=0$.
Hence the components of transformations corresponding to the equation variables have the simple general form 
\begin{equation}\label{EqvcKdVEquivGroup}
\tilde t=\alpha(t),\quad
\tilde x=\beta(t)x+\gamma(t),\quad
\tilde u=\theta(t)u+\varphi(t)x+\psi(t),\quad 
\end{equation}
where $\alpha$, $\beta$, $\gamma$, $\theta$, $\varphi$ and $\psi$ run through the set of smooth functions of~$t$, $\alpha\beta\theta\ne0$. 
The arbitrary elements of~\eqref{EqvcKdV} are transformed as follows
\begin{gather*}
\tilde f=\frac{\beta}{\alpha_t\theta}f, \quad
\tilde g=\frac{\beta^3}{\alpha_t}g, \quad
\tilde h=\frac1{\alpha_t}\left(h-\frac\varphi\theta f-\frac{\theta_t}\theta\right), \\
\tilde q=\frac1{\alpha_t}\left(q-\frac\varphi\theta f+\frac{\beta_t}\beta\right), \quad
\tilde p=\frac1{\alpha_t}\left(\beta p-\gamma q+\frac{\gamma\varphi-\beta\psi}\theta f+\gamma_t-\gamma\frac{\beta_t}\beta\right), \\
\tilde k=\frac1{\alpha_t\beta}\left(\theta k-\varphi(h+q)+\frac{\varphi^2}\theta f-\varphi_t+\varphi\frac{\theta_t}\theta\right), \\
\tilde l=\frac1{\alpha_t}\left(\theta l-\frac{\gamma}{\beta}\left(\theta k-\varphi(h+q)+\frac{\varphi^2}\theta f-
\varphi_t+\varphi\frac{\theta_t}\theta\right)-\psi\left(h-\frac\varphi\theta f-\frac{\theta_t}\theta\right)-\varphi p-\psi_t\right).
\end{gather*}
Any equation from class~\eqref{EqvcKdV} can be reduced by point transformations 
to the form~\eqref{EqZhang2007} with $g=1$ and forms~\eqref{EqZhang2007Reduced} and~\eqref{EqZhang2007Mapped}.
The necessary and sufficient condition of similarity of such equations to the standard KdV equation is 
\begin{equation}\label{EqvcKdVEquivToKdV}
s_t=2gs^2-3qs+\frac fgk, \quad\mbox{where}\quad s:=\frac{2q-h}g+\frac{f_tg-fg_t}{fg^2}.
\end{equation}

The second subclass consists of the variable-coefficient mKdV equations
\begin{equation}\label{EqvcmKdV}
u_t+f(t)u^2u_x+g(t)u_{xxx}+h(t)u+(p(t)+q(t)x)u_x+k(t)uu_x+l(t)=0,
\end{equation}
where all the parameters are arbitrary smooth functions of~$t$, $fg\ne0$. 
This subclass is also normalized. 
Its equivalence group (in terms of the arbitrary element~$H$) is singled out from~$G^{\sim}_4$ by the condition $\Phi_x=0$,
i.e., the components of transformations corresponding to the equation variables are of the form~\eqref{EqvcKdVEquivGroup} with $\varphi=0$,
where $\alpha$, $\beta$, $\gamma$, $\theta$ and $\psi$ run through the set of smooth functions of~$t$, $\alpha\beta\theta\ne0$. 
The arbitrary elements of~\eqref{EqvcmKdV} are transformed by the formulas
\begin{gather*}
\tilde f=\frac{\beta}{\alpha_t\theta^2}f, \quad
\tilde g=\frac{\beta^3}{\alpha_t}g, \quad
\tilde h=\frac1{\alpha_t}\left(h-\frac{\theta_t}\theta\right), \\
\tilde q=\frac1{\alpha_t}\left(q+\frac{\beta_t}\beta\right), \quad
\tilde p=\frac1{\alpha_t}\left(\beta p-\gamma q+\beta\frac{\psi^2}{\theta^2} f-\beta\frac\psi\theta k+\gamma_t-\gamma\frac{\beta_t}\beta\right), \\
\tilde k=\frac\beta{\alpha_t\theta}\left(k-2\frac\psi\theta f\right), \quad
\tilde l=\frac1{\alpha_t}\left(\theta l-\psi h-\psi_t+\psi\frac{\theta_t}\theta\right)
\end{gather*}
Five of the arbitrary elements can be gauged to simple constant values. 
For example, it is possible to set $g=1$ and $h=p=q=l=0$.
An equation of form~\eqref{EqvcmKdV} is similar to the standard mKdV equation if and only if 
\[
2h-2q=\frac{f_t}f-\frac{g_t}g, \quad
2lf=k_t+kh-k\frac{f_t}f.
\]

In Section~\ref{SectionOnExamplesOnSimilarityOfEqsAndTheirSolutions} we present different examples on similarity of equations from 
classes~\eqref{EqvcKdV} and~\eqref{EqvcmKdV} to the standard KdV and mKdV equations, respectively.

Note that hierarchies of normalized classes in the same superclass, which are different from the constructed hierarchy, 
can be easily obtained according as purposes of investigation.

\section{Group analysis of a class of variable-coefficient KdV equations}
\label{SectionOnGroupAnalysisOfClassOfvcKdVeqs}

We study a class of variable-coefficient KdV equations in detail 
since this creates a basis for the consideration of four examples in Section~\ref{SectionOnParticularSimilarity}.
The general form of these equations is 
\begin{equation}\label{EqZhang2007}
u_t+f(t)uu_x+g(t)u_{xxx}=0,
\end{equation}
where $f$ and $g$ are arbitrary (smooth) functions of~$t$, $fg\ne0$. 

As~shown in Section~\ref{SectionONAdmissiblePointTransInClassesOfGenKdVEqs}, for the general values of the parameter-functions (arbitrary elements) $f$ and $g$ 
equation~\eqref{EqZhang2007} is not equivalent to the standard KdV equation up to point transformations
but at least one of the parameters ($f$ or $g$) can be set equal to 1 using a point transformation. 
(The equations of the form~\eqref{EqZhang2007} with $g=1$ are called the transitional KdV equations~\cite{Calogero&Degasperis1982}.)
Thus, after the transformation 
\begin{equation}\label{EqZhang2007Trans}
\tilde t=\int f(t)\,dt,\quad
\tilde x=x,\quad
\tilde u=u
\end{equation}
equation~\eqref{EqZhang2007} takes the same form with $\tilde f(\tilde t)=1$ and $\tilde g(\tilde t)=g(t)/f(t)$.
Therefore, without lost of generality we can consider the class of equations 
\begin{equation}\label{EqZhang2007Reduced}
u_t+uu_x+g(t)u_{xxx}=0,
\end{equation}
where $g$ is an arbitrary (smooth) nonvanishing function of~$t$, 
since this class is the image of class~\eqref{EqZhang2007} under the mapping generated by the family of point transformations~\eqref{EqZhang2007Trans}. 
(See~\cite{Popovych&Kunzinger&Eshraghi2009,Vaneeva&Popovych&Sophocleous2009} for related definitions.) 

We carried out the exhaustive group classification of class~\eqref{EqZhang2007Reduced}. 
Note that Lie symmetries and admissible (``allowed'') transformations of the wider class of equations 
having the form 
\begin{equation}\label{EqWinternitz&Gazeau1992}
u_t+f(t,x)uu_x+g(t,x)u_{xxx}=0 
\end{equation}
were investigated in \cite{Winternitz&Gazeau1992}.
The separate consideration of subclass~\eqref{EqZhang2007Reduced} is additionally justified by the fact that 
it has nicer transformational properties than the superclasses~\eqref{EqZhang2007} and, especially,~\eqref{EqWinternitz&Gazeau1992}.

The equivalence group~$G^\sim$ of class~\eqref{EqZhang2007Reduced} consists of the transformations 
\begin{gather}\label{EqZhang2007ReducedEquivGroup}
\begin{split}
&\tilde t=\frac{at+b}{ct+d},\quad
\tilde x=\frac{e_2x+e_1t+e_0}{ct+d},\\[.5ex]
&\tilde u=\frac{e_2(ct+d)u-e_2cx-e_0c+e_1d}\varepsilon,\quad
\tilde g=\frac{e_2{}^3}{ct+d}\frac g\varepsilon,
\end{split}
\end{gather}
where $a$, $b$, $c$, $d$, $e_0$, $e_1$ and $e_2$ are arbitrary constants, $\varepsilon=ad-bc\ne0$,
and without loss of generality we can assume that $\varepsilon=\pm1$.
The group $G^\sim$ is isomorphic to the matrix group 
\[
\left\{ \left.
\left(\begin{array}{ccc}
e_2&e_1&e_0\\0&a&b\\0&c&d 
\end{array}\right) 
\ \right|\ \ e_2\ne0,\ ad-bc=\pm1\ 
\right\}.             
\]

Equations from class~\eqref{EqZhang2007Reduced} are similar only if they are $G^\sim$-equivalent. 
Moreover, all admissible transformations in this class are generated by transformations from $G^\sim$, 
i.e., the class~\eqref{EqZhang2007Reduced} is normalized in the usual sense. 
(The initial class~\eqref{EqZhang2007} is normalized only with respect to the extended generalized equivalence group 
and class~\eqref{EqWinternitz&Gazeau1992} possesses no normalization properties.) 
This implies the following claim:
\emph{An equation of form~\eqref{EqZhang2007Reduced} is similar to the KdV equation if and only if $g_{tt}=0$. 
Any transformation realizing the similarity belongs to~$G^\sim$.}
Therefore, an equation of form~\eqref{EqZhang2007} is reduced to the KdV equation by a point transformation if and only if 
\begin{equation}\label{EqCondOnEquivOfZhang2007ToKdV}
g(t)=f(t)\left(c_1\int f(t)\,dt+c_0\right), 
\end{equation}
where $c_0$ and $c_1$ are constants, $(c_0,c_1)\ne(0,0)$~\cite{Grimshaw1979},
that well agrees with Theorem~3 of~\cite{Winternitz&Gazeau1992} (cf.\ also equation~\eqref{EqvcKdVEquivToKdV}). 
Equation~\eqref{EqCondOnEquivOfZhang2007ToKdV} coincides with the constraint on arbitrary elements 
of the equations from class~\eqref{EqZhang2007} which have the Painlev\'e property~\cite{Joshi1987}.

Denote by $A^g$ the maximal Lie invariance algebra of the equation of form~\eqref{EqZhang2007Reduced} 
with the fixed value~$g$ of the arbitrary element. 
It follows from the infinitesimal invariance criterion \cite{Olver1986,Ovsiannikov1982}
that the coefficients of any vector field $Q=\tau(t,x,u)\p_t+\xi(t,x,u)\p_x+\eta(t,x,u)\p_u$ from $A^g$ satisfy 
the system of determining equations 
\begin{gather}
\tau_x=\tau_u=\xi_u=\eta_{uu}=\xi_{xx}=\eta_{xu}=0,\quad 
\eta=(\xi_x-\tau_t)u+\xi_t,\quad 
\eta_t+\eta_xu=0,\label{EqZhang2007DetEq1}\\
\tau g_t=(3\xi_x-\tau_t)g.\nonumber
\end{gather}
Integrating equations~\eqref{EqZhang2007DetEq1} gives the following expressions for the coefficients of~$Q$:
\begin{gather*}
\tau=c_2t^2+c_1t+c_0,\\
\xi=(c_2t+c_3)x+c_4t+c_5,\\
\eta=(-c_2t+c_3-c_1)u+c_2x+c_4,
\end{gather*}
involving only arbitrary constants $c_0$, \ldots, $c_5$, i.e. for any~$g\ne0$ the algebra~$A^g$ is finite dimensional. 
Then the last determining equation leads to the unique \emph{classifying condition}
\[
(c_2t^2+c_1t+c_0)g_t=(c_2t+3c_3-c_1)g.
\]
The group classification of class~\eqref{EqZhang2007Reduced} is equivalent to the integration of the classifying condition 
up to the $G^\sim$-equivalence. 
The transformation~\eqref{EqZhang2007ReducedEquivGroup} acts on the coefficients of the equation 
$(\alpha t^2+\beta t+\gamma)g_t=(\alpha t+\delta)g$ in the following way:
\begin{gather*}
\alpha=a^2\tilde\alpha+ac\tilde\beta+c^2\tilde\gamma,\\
\beta=2ab\tilde\alpha+(ad+bc)\tilde\beta+2cd\tilde\gamma,\\
\gamma=b^2\tilde\alpha+bd\tilde\beta+d^2\tilde\gamma,\\
\delta=ab\tilde\alpha+bc\tilde\beta+cd\tilde\gamma+\varepsilon\tilde\delta.
\end{gather*}

The kernel $A^\cap=\cap_{g\ne0}A^g$ of the maximal Lie invariance algebras of equations from class~\eqref{EqZhang2007Reduced} 
is $A^\cap=\langle\p_x,\,t\p_x+\p_u\rangle$. 
All $G^\sim$-inequivalent cases of Lie symmetry extension are exhausted by the following:

\medskip

\noindent
1. $\displaystyle g=\exp\left(\int \frac{\alpha t+\delta}{\alpha t^2+\beta t+\gamma}dt\right)$, 
$(\alpha,\delta)\ne(0,0)$, $(\alpha,\beta,\gamma)\ne(0,0,0)$: \\
$A^g=\langle\p_x,\,t\p_x+\p_u,\, 3(\alpha t^2+\beta t+\gamma)\p_t+(3\alpha t+\delta+\beta)x\p_x+((-3\alpha t+\delta-2\beta)u+3\alpha x)\p_u\rangle$.

\medskip

\noindent
Up to $G^\sim$-equivalence this case is partitioned into the three inequivalent subcases:

\medskip

$g=e^t$: $A^g=\langle\p_x,\,t\p_x+\p_u,\, 3\p_t+x\p_x+u\p_u\rangle$ 

($\alpha=\beta=0$, $\delta/\gamma$ is scaled to~1);

\medskip

$g=t^\mu$, $\mu\geqslant1/2$, $\mu\ne1$: $A^g=\langle\p_x,\,t\p_x+\p_u,\, 3t\p_t+(\mu+1)x\p_x+(\mu-2)u\p_u\rangle$

($\alpha=\gamma=0$, $\mu:=\delta/\beta$); 

\medskip

$g=e^{\delta\arctan t}\sqrt{t^2+1}$: $A^g=\langle\p_x,\,t\p_x+\p_u,\, 3(t^2+1)\p_t+(3t+\delta)x\p_x+((-3t+\delta)u+3x)\p_u\rangle$

($\alpha=\gamma=1$, $\beta=0$); 

\medskip

\noindent
2. $g=1$: 
$A^g=\langle\p_x,\,t\p_x+\p_u,\,\partial_t,\,3t\partial_t+x\partial_x-2u\partial_u\rangle$.

\medskip

The presented group classification gives all inequivalent values of~$g$ for which 
the classical method of Lie reduction can be effectively used. 

The class~\eqref{EqZhang2007} can be also mapped to the class 
\begin{equation}\label{EqZhang2007Mapped}
\tilde u_{\tilde t}+\tilde u\tilde u_{\tilde x}+\tilde u_{\tilde x\tilde x\tilde x}+h(\tilde t)\tilde u=0
\end{equation}
by the family of point transformations
\[
\tilde t=\int g(t)\,dt,\quad
\tilde x=x,\quad
\tilde u=\frac fgu.
\]
The arbitrary element~$h$ of the mapped class is expressed via the arbitrary elements~$f$ and~$g$ 
in the following way:
\[
h(\tilde t)=\frac{f(t)g_t(t)-f_t(t)g(t)}{f(t)(g(t))^2}.
\]
Thus, the equation~\eqref{EqZhang2007Reduced} with $g=t$ is then mapped to the cylindrical KdV equation ($h=(2t)^{-1}$) 
whose similarity to the standard KdV equation is known for a long time~\cite{Lugovtsov&Lugovtsov1969}. 
Analogously, the value $g=e^t$ corresponds to the spherical KdV equation ($h=t^{-1}$) which is not integrable.

\section{Examples on similarity of equations and their solutions}
\label{SectionOnExamplesOnSimilarityOfEqsAndTheirSolutions}

As shown in Section~\ref{SectionONAdmissiblePointTransInClassesOfGenKdVEqs},
to establish similarity of equations in all the presented examples it is sufficient to apply 
point transformations of the simple general form~\eqref{EqvcKdVEquivGroup}.
In some of the examples presented we change the notations of arbitrary elements in equations in order to unify the consideration. 
Thus, the arbitrary elements $f$, $g$, $h$, $k$, $l$, $p$, $q$ and $s$ run through the set of smooth functions of~$t$ 
satisfying certain conditions of nonvanishing, which are explicitly indicated in the examples. 
We also explicitly indicate necessary conditions of nonsingularity for arbitrary elements 
even if they were not given in the corresponding commented papers. 
We should like to emphasize that correctness of solutions given in commented papers is not discussed here. 
Other common errors of different kinds described in~\cite{Kudryashov2009} were made in the most of commented papers.

\subsection{Similarity of entire classes}
\label{SectionOnExamplesOnCompleteSimilarity}

In each example collected in this section, all equations of the class under consideration 
are similar to a single well-studied equation (either the KdV or mKdV ones). 
This means that the arbitrary elements (constant or functional parameters) 
introduced in order to generalize the classical equations to such classes are completely needless.

\begin{example} 
In \cite{Zhang&Tong&Wang2008} the so-called ``generalized $(G'/G)$-expansion method'' 
was applied to finding exact solutions of ``the mKdV equations with variable coefficients'' having the form 
\begin{equation}\label{EqZhang&Tong&Wang2008}
u_t=K_0(t)(u_{xxx}-6u^2u_x)+4K_1(t)u_x-h(t)(u+xu_x),
\end{equation}
where $K_0$, $K_1$ and $h$ are arbitrary (smooth) functions of~$t$, $K_0\ne0$. 
The same class of equations was investigated, e.g., in~\cite{Dai&Zhu&Zhang2006},~\cite{Hong2009} and~\cite{Triki&Wazwaz2009}
using so-called ``the variable-coefficient generalized projected Ricatti equation expansion method'',
``generalized Jacobi elliptic functions expansion method'' and ``sub-ODE method'', respectively. 

At the same time, any equation of form~\eqref{EqZhang&Tong&Wang2008} is equivalent to the standard mKdV equation 
with respect to a point transformation.  
Indeed, consider the point transformation 
\begin{equation}\label{EqZhang&Tong&Wang2008Trans}
\tilde t=\alpha(t),\quad
\tilde x=\beta(t)x+\gamma(t),\quad
\tilde u=\varepsilon\frac u{\beta(t)},
\end{equation}
where $\alpha$, $\beta$ and $\gamma$ are smooth functions of~$t$, which should be additionally specified, $\alpha_t\beta\ne0$ and $\varepsilon\in\{-1;1\}$.
It maps equation~\eqref{EqZhang&Tong&Wang2008} to the equation
\[
\alpha_t\tilde u_{\tilde t}+\frac{\beta_t}\beta(\tilde u+\beta x\tilde u_{\tilde x})+\gamma_t\tilde u_{\tilde x}
=K_0\beta^3(\tilde u_{\tilde x\tilde x\tilde x}-6\tilde u^2\tilde u_{\tilde x})
+4K_1\beta\tilde u_{\tilde x}-h(\tilde u+\beta x\tilde u_{\tilde x}).
\] 
If, e.g., $\varepsilon=1$ and the functions $\alpha$, $\beta$ and $\gamma$ satisfy the system 
\[
\alpha_t=-K_0\beta^3,\quad
\beta_t=-h\beta,\quad
\gamma_t=4K_1\beta,
\]
the transformed equation is nothing but the standard mKdV equation 
$\tilde u_{\tilde t}-6\tilde u^2\tilde u_{\tilde x}+\tilde u_{\tilde x\tilde x\tilde x}=0$.
In other words, the function $u=u(t,x)$ satisfies equation~\eqref{EqZhang&Tong&Wang2008} if and only if 
there exists a solution $\tilde u=\tilde u(\tilde t,\tilde x)$ of the standard mKdV equation such that 
\[
u=\beta\tilde u(\alpha,\beta x+\gamma), 
\]
where 
$
\alpha=-\int K_0e^{-3\int h\,dt}dt,\ 
\beta=e^{-\int h\,dt},\ 
\gamma=4\int K_1e^{-\int h\,dt}dt.
$

Note additionally that the usual equivalence group \cite{Popovych&Kunzinger&Eshraghi2009,Ovsiannikov1982} of class~\eqref{EqZhang&Tong&Wang2008} consists of 
the transformations of form~\eqref{EqZhang&Tong&Wang2008Trans} 
prolonged to the arbitrary elements $K_0$, $K_1$ and $h$ by the formulas 
\[
\tilde K_0=\frac{\beta^3}{\alpha_t}K_0,\quad
\tilde K_1=\frac1{\alpha_t}\left(\beta K_1-\frac{\gamma_t}4+\frac\gamma4\left(h+\frac{\beta_t}\beta\right)\right),\quad
\tilde h=\frac1{\alpha_t}\left(h+\frac{\beta_t}\beta\right). 
\] 
Moreover, if two fixed equations from the class~\eqref{EqZhang&Tong&Wang2008} are connected via a point transformation 
then this transformation has the form~\eqref{EqZhang&Tong&Wang2008Trans}. 
This means that the class~\eqref{EqZhang&Tong&Wang2008} is a normalized class of differential equations 
which is a single orbit of its equivalence group, generated from the standard mKdV equation. 
\end{example}

\begin{example}\label{ExampleWu&Xia2008AndWang&Wang&Zhou2002}
Some traveling wave solutions of the ``compound/combined KdV--mKdV equation''
\begin{equation}\label{EqWu&Xia2008}
u_t+(\alpha+\beta u)uu_x+\gamma u_{xxx}= 0,
\end{equation}
where $\alpha$, $\beta$ and $\gamma$ are real constants, $\beta\gamma\ne0$, were constructed 
in~\cite{Ebaid2007,Inan&Kaya2007,Sirendaoreji2006,Wu&Xia2008,Yomba2005,Zhang&Xia2008,Zhao&Zhi&Yu&Zhang2006}. 
In fact, this equation is called the Gardner equation ($\alpha$ should be scaled to a standard value) 
and is obviously similar to the mKdV equation 
$\tilde u_{\tilde t}+\varepsilon\tilde u^2\tilde u_{\tilde x}+\tilde u_{\tilde x\tilde x\tilde x}= 0$, where $\varepsilon=\sign(\beta\gamma)$, 
with respect to the point transformation 
\[
\tilde t=\gamma t,\quad
\tilde x=x+\frac{\alpha^2}{4\beta}t,\quad
\tilde u=\sqrt{\left|\frac\beta\gamma\right|}\left(u+\frac\alpha{2\beta}\right)
\]
which is well known for a long time~\cite{Miura1968}.
Therefore, each solution of equation~\eqref{EqWu&Xia2008} is represented in the form 
\[
u(t,x)=\sqrt{\left|\frac\gamma\beta\right|}\ \tilde u\left(\gamma t,x+\frac{\alpha^2}{4\beta}t\right)-\frac\alpha{2\beta},
\]
where $\tilde u$ is a solution of the mKdV equation, 
and for any solution~$\tilde u$ of the mKdV equation this representation gives a solution of~\eqref{EqWu&Xia2008}.

The other close class of equations 
\begin{equation}\label{EqYusufoglu2008}
u_t+u_x+\alpha u^2u_x+u_{xxx}=0,
\end{equation}
where $\alpha$ runs through the set of real nonvanishing constants, was considered in \cite{Yusufoglu2008}. 
Only specific traveling wave solutions were found using the so-called ``Exp-function method''. 
The equations of form~\eqref{EqYusufoglu2008} were groundlessly called the modified Benjamin--Bona--Mahony equations. 
In fact the Benjamin--Bona--Mahony equation $u_t+u_x+uu_x-u_{xxt}=0$ has no, at least, direct relation with class~\eqref{EqYusufoglu2008}. 
At the same time, any equation of form~\eqref{EqYusufoglu2008} is reduced by the trivial point transformation 
\[
\tilde t=t,\quad
\tilde x=x-t,\quad
\tilde u=\sqrt{|\alpha|}u
\]
to the mKdV equation 
$\tilde u_{\tilde t}+\varepsilon\tilde u^2\tilde u_{\tilde x}+\tilde u_{\tilde x\tilde x\tilde x}= 0$, where $\varepsilon=\sign\alpha$.
Equation~\eqref{EqYusufoglu2008} with $\alpha=1$ was also investigated using ``extended F-expansion method'' in~\cite{Liu&Yang2004}.
\end{example} 

\begin{example}
The authors of~\cite{Sabry&El-Taibany2008} apply ``generalized expansion method''
to find exact solutions of generalized KdV equations with variable coefficients, which have the form 
\begin{equation}\label{Sabry&El-Taibany2008}
u_t+g(t)(6uu_x+u_{xxx})+6f(t)g(t)u=x(f_t(t)+12g(t)f^2(t))+M(t)
\end{equation}
with  $g\ne0$.
The whole class~\eqref{Sabry&El-Taibany2008} is mapped to the KdV equation 
$\tilde u_{\tilde t}+6\tilde u\tilde u_{\tilde x}+\tilde u_{\tilde x\tilde x\tilde x}= 0$ by the family of point transformations 
\[
\tilde t=\int g\gamma^3\,dt,\quad
\tilde x=\gamma x-6\int g\,\gamma^3\beta\,dt,\quad
\tilde u=\frac{u-fx}{\gamma^2}-\beta,
\]
where $\gamma=e^{-6\int fg\, dt}$ and $\beta=\int M\gamma^{-2}\,dt$. 
This means that the function $u=u(t,x)$ satisfies an equation of form~\eqref{Sabry&El-Taibany2008} if and only if 
it is represented via  a solution $\tilde u$ of the KdV equation by the expression 
\[
u=\gamma^2\,\tilde u\left(\int g\gamma^3\,dt,\gamma x-6\int g\,\gamma^3\beta\,dt\right)+fx+\gamma^2\beta.
\]

An auto-B\"acklund transformation and exact solutions of equations from class~\eqref{Sabry&El-Taibany2008} with $M=0$
were considered in~\cite{Wang&Wang&Zhou2002}.
All related results also can be reproduced from corresponding results for the KdV equation using the above point transformations, where $M=0$.  

The subclass of the equations~\eqref{Sabry&El-Taibany2008} with $g=1$ and $M=0$ arose in~\cite{Chou1987}, 
where symmetry properties of such equations were studied. 
It was also mentioned in~\cite{Chou1987} that obtained results can be extended to the equations of the more general form 
\begin{equation}\label{EqChou1987}
u_t+6uu_x+u_{xxx}+6f(t)u=x(f_t(t)+12f^2(t))+h_t(t)+12f(t)h(t).
\end{equation}
The family of point transformations mapping class~\eqref{EqChou1987} to the KdV equation 
consists of the transformations 
\[
\tilde t=\int \gamma^3\,dt,\quad
\tilde x=\gamma x-6\int h\gamma\,dt,\quad
\tilde u=\frac{u-fx-h}{\gamma^2},
\]
where $\gamma=e^{-6\int f\, dt}$. 
\end{example}

\begin{example} 
Using the so-called ``Jacobi elliptic function expansion method'',
some new soliton-like solutions were obtained in~\cite{Zhao&Tang&Wang2005}
for the KdV-like equations with variable coefficients, 
which have the general form 
\begin{equation}\label{EqZhao&Tang&Wang2005}
u_t+f(t)(xu_x+2u)+g(t)(u_{xxx}+auu_x)+h(t)u_x=0.
\end{equation}
Here $a$ is an arbitrary constant 
and $ag\ne0$.
Any equation from class~\eqref{EqZhao&Tang&Wang2005} is similar 
to the standard KdV equation $\tilde u_{\tilde t}+\tilde u\tilde u_{\tilde x}+\tilde u_{\tilde x\tilde x\tilde x}=0$
with respect to the point transformation  
\begin{equation}\label{EqZhao&Tang&Wang2005Trans}
\tilde t=\alpha(t),\quad
\tilde x=\beta(t)x+\gamma(t),\quad
\tilde u=\frac a{(\beta(t))^2}u,
\end{equation}
where the parameter-functions $\alpha$, $\beta$ and $\gamma$ 
are expressed via arbitrary elements of class~\eqref{EqZhao&Tang&Wang2005} in the following way:
\begin{gather*}
\beta=e^{-\int f\,dt},\quad
\alpha=\int g\beta^3\,dt,\quad
\gamma=-\int h\beta\,dt.
\end{gather*}
\end{example}

\subsection{Particular similarity}
\label{SectionOnParticularSimilarity}

Although the classes of evolution equations from examples of this section are not similar in whole to classical equations, 
they also involve needless arbitrary elements. 
Canceling such arbitrary elements with point transformations leads to essential simplification of calculations. 
Moreover, in all the papers commented here results were in fact derived only for specific values of arbitrary elements 
for which the corresponding equations are similar to either the KdV or mKdV equations.

\begin{example}\label{ExampleZhang2007}
Class~\eqref{EqZhang2007} of KdV-like equations with coefficients depending on~$t$ 
was considered in \cite{Zhang2007,Zhang&Zhang2009} using the so-called ``Exp-function method''. 
(See also references in~\cite{Zhang2007,Zhang&Zhang2009} on a number of papers 
where some particular results were obtained for the same class in similar frameworks.)
Auto-B\"acklund transformations and some analytical solutions for equations from class~\eqref{EqZhang2007} 
were considered in \cite{Fan2002,Hong&Jung1999,Wang&Wang2001}.
In fact, certain results are obtained in the above papers only for the particular case $g/f=\const$ of~\eqref{EqCondOnEquivOfZhang2007ToKdV}
which is obviously reduced to the KdV equation~\cite{Grimshaw1979}. 
The corresponding point transformations combined with the known wide families of exact solutions of the KdV equation 
give at once much more knowledge about exact solutions of equations from class~\eqref{EqZhang2007} 
than that presented in~\cite{Zhang2007,Zhang&Zhang2009} and previous papers on the subject. 
In~\cite{Wu2009} Lax pairs were constructed only for equations whose arbitrary elements satisfy condition~\eqref{EqCondOnEquivOfZhang2007ToKdV}.
This is why the results of~\cite{Wu2009} are obvious consequences of classical results for the KdV equation. 
See the discussion on Lax pairs in the conclusion. 
\end{example}

\begin{example}
In~\cite{Sabry&Zahran&Fan2004} ``generalized variable coefficients KdV equations'' of the form   
\begin{equation}\label{EqSabry&Zahran&Fan2004}
u_t+f(t)uu_x+g(t)u_{xxx}+l(t)=0
\end{equation}
with $fg\ne0$ were considered using a ``new generalized expansion method''.
%
The coefficient~$l$ can be made equal to zero by a point transformation. 
More precisely, any equation from class~\eqref{EqSabry&Zahran&Fan2004} is mapped to the equation from class~\eqref{EqZhang2007} with 
the same values of the arbitrary elements~$f$ and~$g$ by the transformation 
\[
\tilde t=t,\quad
\tilde x=x+\gamma(t),\quad
\tilde u=u+\psi(t),
\]
where $\gamma_t=f\psi$ and $\psi_t=l$. 
Since class~\eqref{EqSabry&Zahran&Fan2004} is a preimage of class~\eqref{EqZhang2007} and, therefore, class~\eqref{EqZhang2007Reduced} 
with respect to mapping generated by families of point transformations, 
it is needless to construct exact solutions of equations of form~\eqref{EqSabry&Zahran&Fan2004} with nonzero values of~$l$. 

Moreover, exact solutions of~\eqref{EqSabry&Zahran&Fan2004} were constructed in~\cite{Sabry&Zahran&Fan2004} only in the case 
\[
3gf_t{}^2-3ff_tg_t-fgf_{tt}+f^2g_{tt}=0
\] 
when the corresponding equation~\eqref{EqSabry&Zahran&Fan2004} is obviously similar to the KdV equation 
in view of that the above equation in~$f$ and~$g$ is equivalent to condition~\eqref{EqCondOnEquivOfZhang2007ToKdV}. 
\end{example}

\begin{example}
At first sight the class of ``KdV equations with variable coefficients''
\begin{equation}\label{EqZhang2008}
u_t+f(t)uu_x+g(t)u_{xxx}+h(t)u+(p(t)+q(t)x)u_x=0
\end{equation}
considered in~\cite{Liu&Yang2004,Wei2009,Zhang2008} looks quite complicated 
since it involves the five (!)\ arbitrary smooth functions $f$, $g$, $h$, $p$ and $q$ of~$t$, where $fg\ne0$. 

The form~\eqref{EqZhang2008} was partially simplified in~\cite{Zhang2008} in the beginning of consideration.
Namely, the coefficient~$h$ was gauged to zero by a point transformation. 
In fact only one arbitrary element among $f$, $g$, $h$, $p$ and $q$ is essential. 
The other ones can be set equal to standard fixed values. 
Indeed, class~\eqref{EqZhang2008} is mapped to class~\eqref{EqZhang2007} by the family of point transformations 
\[
\tilde t=t,\quad
\tilde x=\beta(t)x+\gamma(t),\quad
\tilde u=\theta(t)u,
\]
where $\theta=e^{\int h\,dt}$, $\beta=e^{-\int q\,dt}$  and $\gamma=-\int p\beta\,dt$. 
The corresponding value of the imaged arbitrary elements are 
$\tilde f=f\beta/\theta$ and $\tilde g=g\beta^3$.
The class~\eqref{EqZhang2007} can be then transformed, e.g., to class~\eqref{EqZhang2007Reduced} or class~\eqref{EqZhang2007Mapped}.

Exact solutions were constructed in~\cite{Wei2009,Zhang2008} only for equations on form~\eqref{EqZhang2008} with arbitrary elements 
constrained by the condition
\[
fe^{\int (h-2q)\,dt}/g=\const
\]
which is written down in terms of the imaged arbitrary elements as $\tilde f/\tilde g=\const$. 
In~\cite{Liu&Yang2004} the constraint $h=2q$ was additionally imposed.
Hence all such equations are similar to the KdV equation (cf. Example~\ref{ExampleZhang2007}) 
and the construction of their exact solutions is needless. 
\end{example}

\begin{example} 
The class of systems of the form 
\begin{equation}\label{EqZhou&Wang&Wang2005}\begin{split}
&u_t+h(t)uu_x+s(t)vv_x+g(t)u_{xxx}=0,\\
&v_t+f(t)uv_x+g(t)v_{xxx}=0
\end{split}\end{equation}
generalizing the Hirota--Satsuma system 
were studied in~\cite{Zhou&Wang&Wang2005} by using so-called ``$F$-expan\-sion method''.
Here $f$, $g$, $h$ and $s$ are arbitrary smooth functions of~$t$ satisfying the conditions
\[
fgs\ne0,\quad g/f=\const,\quad f-h=\sigma^2s,
\] 
where $\sigma$ is a constant. 
The parameter-function~$g$ could be at once set equal to 1 by a simple transformation only of~$t$ (namely, $\tilde t=\int g\,dt$), 
then the transformed element~$f$ would be an arbitrary nonvanishing constant. 

The solution presented in~\cite{Zhou&Wang&Wang2005} satisfies the additional condition $v=-\sigma u$ 
under which the system~\eqref{EqZhou&Wang&Wang2005} is reduced, in view of the condition $f-h=\sigma^2s$, 
to the single equation~\eqref{EqZhang2007} which is similar to the KdV equation since $g/f=\const$ 
(cf. Example~\ref{ExampleZhang2007}).  
As a result, the solutions of~\eqref{EqZhou&Wang&Wang2005} found in~\cite{Zhou&Wang&Wang2005} in fact can be easily derived 
from simplest traveling wave solutions of the KdV equation. 
\end{example}

\begin{example}\label{ExampleYan2008AndXu&Meng&Gao&Wen2009} 
In~\cite{Yan2008} some exact solutions were constructed for variable coefficient mKdV equations of the general form 
\begin{equation}\label{EqYan2008}
u_t+f(t)u^2u_x+g(t)u_{xxx}+p(t)u_x=0,
\end{equation}
where 
$fg\ne0$  
and additionally the condition $g/f=\const$ should be satisfied. 
At the same time, any equation of form~\eqref{EqYan2008}, whose coefficients are constrained by this condition, 
is mapped by the point transformation 
\[
\tilde t=\int g(t)\,dt,\quad
\tilde x=x-\int p(t)\,dt,\quad
\tilde u=\sqrt{|f/g|}u, 
\]
to the mKdV equation 
$\tilde u_{\tilde t}+\varepsilon\tilde u^2\tilde u_{\tilde x}+\tilde u_{\tilde x\tilde x\tilde x}= 0$, where $\varepsilon=\sign(fg)$.
Without additional supposition, an equation from class~\eqref{EqYan2008} is reduced by point transformations to 
an equation from the same class in which, e.g., $p=0$ and $f=1$. 
Another possibility is to gauge the coefficients~$p$ and $g$ to the values~$0$ and~$1$, respectively.

The more general class of equations 
\begin{equation}\label{EqXu&Meng&Gao&Wen2009}
u_t+f(t)u^2u_x+k(t)uu_x+g(t)u_{xxx}+p(t)u_x+h(t)u=0
\end{equation}
with $fg\ne0$ was considered in~\cite{Xu&Meng&Gao&Wen2009} within the framework of Hirota's approach.
Only the equations of form~\eqref{EqXu&Meng&Gao&Wen2009} 
whose arbitrary elements satisfy the additional constraint $e^{-2\int h\,dt}f/g=\const$ were represented in bilinear forms. 
One- and multi-solitary-wave solutions of such equations were found after imposing the one 
more additional constraint $e^{-\int h\,dt}k/g=\const$. 
(Under the construction of multi-solitary-wave solutions this constraint was used implicitly.)

Exact solutions for the equations  from  class~\eqref{EqXu&Meng&Gao&Wen2009} were also constructed in~\cite{Wei2009} 
using the so-called ``extended tanh method''.  
The solutions also were obtained only for those equations which coefficients satisfy the system of both above constraints.

Any equation from class~\eqref{EqXu&Meng&Gao&Wen2009} can be mapped to an equation from the same class, 
in which $\tilde p=0$, $\tilde h=0$ and $\tilde g=1$.
The corresponding transformation of the variables and other arbitrary elements is 
\[
\tilde t=\int g(t)\,dt,\quad
\tilde x=x-\int p(t)\,dt,\quad
\tilde u=e^{\int h\,dt}u,\quad 
\tilde f=\frac{e^{-2\int h\,dt}}{g}f,\quad 
\tilde k=\frac{e^{-\int h\,dt}}{g}k. 
\]
Then the above constraints take the form $\tilde f=\const$ and $\tilde k=\const$, respectively. 
If the arbitrary element~$\tilde f$ is constant, it can be set to~$\pm1$ by scaling~$u$. 
If we additionally have $\tilde k=\const$, this parameter can be made to vanish by a simple transformation 
(cf.\ Example~\ref{ExampleWu&Xia2008AndWang&Wang&Zhou2002}).
Therefore, bilinear forms were given in~\cite{Xu&Meng&Gao&Wen2009} only for equations similar to the equations 
\[u_t+\varepsilon u^2u_x+k(t)uu_x+u_{xxx}=0.\] 
Moreover, in~\cite{Wei2009,Xu&Meng&Gao&Wen2009} exact solutions were found only for equations similar to the mKdV equation!

Note that the above constraints as well as canonical forms of the corresponding equations up to point transformations
had already been presented, e.g., in~\cite{Li&Xu&Meng&Zhang&Zhang&Tian2007} 
(cf. also Section~\ref{SectionONAdmissiblePointTransInClassesOfGenKdVEqs}).
\end{example}

\begin{example} 
Lie symmetries and similarity solutions of generalized KdV equations of the general form 
\begin{equation}\label{EqSenthilkumaran&Pandiaraja&Vaganan2008}
u_t+u^nu_x+g(t)u_{xxx}+h(t)u=0,
\end{equation}
where $n\in\mathbb Z_+$ and 
$g\ne0$, 
were investigated in~\cite{Senthilkumaran&Pandiaraja&Vaganan2008}.
The parameter-function~$h$ can be set equal to zero by the point transformation
\[
\tilde t=\int \theta^{-n}\,dt,\quad
\tilde x=x,\quad
\tilde u=\theta u,
\]
where $\theta=e^{\int h\,dt}$ and 
the transformed value of the arbitrary element~$g$ is $\tilde g(\tilde t)=g(t)(\theta(t))^n$. 
This means that fixing the arbitrary element~$h$ cannot lead to cases of equations~\eqref{EqSenthilkumaran&Pandiaraja&Vaganan2008} 
with special symmetry properties. 
Therefore, results from~\cite{Senthilkumaran&Pandiaraja&Vaganan2008} on Lie symmetries of equations 
from class~\eqref{EqSenthilkumaran&Pandiaraja&Vaganan2008} are at least incomplete. 
\end{example}

\section{Inaccuracies concerning similarity transformations}
\label{SectionOnInaccuraciesConcerningSimilarityTransformations}

In spite of that the similarity transformations in classes of variable-coefficients equations generalizing classical integrable ones 
are well known~\cite{Grimshaw1979,Hirota1979,Hlavaty1988,Popovych&Kunzinger&Eshraghi2009,Winternitz&Gazeau1992}, 
in a recently published series of papers 
(see, e.g., \cite{Li&Xu&Meng&Zhang&Zhang&Tian2007,Tian&Wei&Zhang&Shan&Gao2006,Tian&Chen&Liu2009} and references therein) 
methods of the construction of similarity transformations 
between different generalized KdV (resp.\ mKdV, resp.\ nonlinear Schr\"odinger etc.) equations were presented 
as quite original. 
Although these results are not really new, the above presentations of them contain certain common inaccuracies 
which can be summed up in the following points:
\begin{itemize}\itemsep=0ex
\item
Restrictions on involving arbitrary elements for which found similarity transformations are correct 
are not indicated in an explicit form.
\item
Equations are reduced to a few canonical forms which are similar to each other.
\item
Wide sets of transformations reducing equations to canonical forms are constructed in cases when it is enough to look for only the simplest one. 
At the same time, no proof on the completeness of the sets is presented. 
\item
Cases of reducibility of equations to canonical forms are partitioned into different subsets although such partition is needless. 
\item
Similarity transformations are looked out only in special sets of points transformations without rigorous arguments.
\end{itemize}
For each of the above points there is a simple justification. 
Thus, equations whose arbitrary elements satisfy different conditions of, e.g., 
vanishing or nonvanishing can be reduced to non-similar canonical forms.   
If two canonical forms~$\mathcal L_1$ and~$\mathcal L_2$ are similar with respect to a point transformation~$\mathcal T_0$
then an equation~$\mathcal L$ is reduced to~$\mathcal L_1$ by a point transformation~$\mathcal T_1$ 
if and only if the same is true for~$\mathcal L_2$ with the transformation~$\mathcal T_0\mathcal T_1$. 
A point transformation~$\mathcal T$ maps an equation~$\mathcal L$ to an equation~$\mathcal L'$ if and only if 
it can be represented in the form $\mathcal T=\mathcal T_0\mathcal S=\mathcal S'\mathcal T_0$, 
where $\mathcal T_0$ is a fixed point transformation from~$\mathcal L$ to~$\mathcal L'$, 
$\mathcal S$ and $\mathcal S'$ are point symmetry transformations of~$\mathcal L$ and~$\mathcal L'$, respectively.
Given a class of differential equations and a canonical form which some of these equations are similar to, 
it is better to present a unified condition which is singled out the entire subclass of equations reducible to the canonical form.  
Most physically important classes of differential equations admit only fiber-preserving point transformations 
satisfying different simple conditions 
(see, e.g., Section~\ref{SectionONAdmissiblePointTransInClassesOfGenKdVEqs} and~\cite{Kingston&Sophocleous1998,Popovych&Kunzinger&Eshraghi2009}).

\begin{example} 
Class~\eqref{EqXu&Meng&Gao&Wen2009} was considered in~\cite{Li&Xu&Meng&Zhang&Zhang&Tian2007} 
without indicating explicit conditions on arbitrary elements. 
It was shown that an equation from this class possesses the Panlev\'e property if and only if the corresponding values 
of the arbitrary elements satisfy the additional constraints $e^{-2\int h\,dt}f/g=\const$ and $e^{-\int h\,dt}q/g=\const$. 
Lax pairs, B\"acklund transformations and simplest solutions of such equations were found. 
Then point transformations which reduce the equations possessing the Panlev\'e property to the Gardner and mKdV equations
were constructed. 

Such consideration has a few inaccuracies. 
The general form~\eqref{EqXu&Meng&Gao&Wen2009} should be simplified from the very beginning (cf.\ Example~\ref{ExampleYan2008AndXu&Meng&Gao&Wen2009}). 
Depending on the sign of $f/g$, an equation of form~\eqref{EqXu&Meng&Gao&Wen2009} with the Panlev\'e property 
is similar either the KdV equation or the mKdV equation with the same sign of the nonlinearity. 
This is why at least the condition $f\ne0$ should be explicitly imposed for the mKdV equation to be a canonical form. 
The equations with the Panlev\'e property were reduced also to the Gardner equation which is obviously similar to the mKdV one 
(cf.\ Example~\ref{ExampleWu&Xia2008AndWang&Wang&Zhou2002}) and, therefore, contains, as a canonical form, a needless arbitrary element.
The unnecessary parameters appeared also in constructed transformations. 

Analogous remarks can be made on \cite{Tian&Wei&Zhang&Shan&Gao2006}, 
where equations of the form~\eqref{EqvcKdV} 
with $k=q=0$ and $fg\ne0$ were considered.  
Under proper additional constraints on arbitrary elements, such equations 
were simultaneously reduced to two canonical forms (the standard and cylindrical KdV equations) which are similar. 
Transformations were a priori assumed to be foliation-preserving although the necessity of this assumption can be easily proved. 
The similarity transformations constructed should be factorized with respect to point symmetry transformations.

In~\cite{Wei2009} it was incorrectly claimed that any equation from class~\eqref{EqZhang2008} 
can be reduced to the cylindrical KdV equation.

The similarity of equations of the form 
\begin{equation}\label{EqGenvcKdV}
u_t+F(t,x)uu_x+G(t,x)u_{xxx}+H(t,x)u+P(t,x)u_x+L(t,x)=0
\end{equation}
with $L=0$ to the KdV equation was considered in~\cite{Tian&Chen&Liu2009}. 
Only foliation-preserving point transformations which are linear in~$u$ were chosen for establishing the similarity, 
and the choice was not justified.  
This approach was groundlessly called ``the direct method''. 
The system of determining equations for components of the transformations was presented in an inaccurate way with no simplification 
and integrated only under a strong additional constraint. 
As a result, all equations the similarity of which to the KdV equation was really studied belong to the class~\eqref{EqvcKdV}. 
At the same time, it can be easily proved (cf.\ Section~\ref{SectionONAdmissiblePointTransInClassesOfGenKdVEqs}) 
that the class~\eqref{EqGenvcKdV} is normalized. 
Its equivalence group consists of the transformations whose components corresponding to the equation variables 
have the form~\eqref{EqGenKdVEquivGroup4}.
The necessary and sufficient conditions of the similarity to the KdV equation (which are analogous to~\eqref{EqvcKdVEquivToKdV})
and the corresponding point transformations can be simply derived 
for both the entire class~\eqref{EqGenvcKdV} and its subclass singled out by the constraint~\mbox{$L=0$}.

\end{example}

{
\renewcommand{\refname}{Commented papers}
\makeatletter
\renewcommand{\@biblabel}[1]{[C#1]} 

\def\@bibitem#1{\item\if@filesw \immediate\write\@auxout
       {\string\bibcite{#1}{C\the\value{\@listctr}}}\fi\ignorespaces}

\makeatother

}


\section{Conclusion}

We have discussed the common error in finding exact solutions of nonlinear differential equations, 
which concerns the similarity of equations. 
The consideration is exemplified with a number of papers on variable-coefficient KdV and mKdV equations.  
There also exist flows of similar papers on the variable-coefficient sin-Gordon and Schr\"odinger equations,  
different generalizations of diffusion equations etc., which should be additionally analyzed. 
Note that a part of such papers concerning nonlinear Schr\"odinger equations have been already examined in 
\cite{Kundu2009,Popovych&Kunzinger&Eshraghi2009},

The similarity can be applied not only for generating new solutions from known ones and simplifying calculations.
It also gives different criteria for testing final results. 
Thus, if an arbitrary element of a class of differential equations can be neglected using a point transformation, 
it should play no crucial role in conditions singling out, from this class, subclasses with special symmetry or 
transformational properties or special cases in finding exact solutions.  

The approach based on similarity is easily extended to different local objects and properties related to differential equations, 
e.g., 
Lie and point symmetries~\cite{Popovych&Kunzinger&Eshraghi2009,Ovsiannikov1982,Vaneeva&Popovych&Sophocleous2009},
conservation laws and potential symmetries~\cite{Popovych&Ivanova2005PETs,Popovych&Ivanova2005,Popovych&Kunzinger&Ivanova2008},
reduction operators (i.e., nonclassical symmetries)~\cite{Popovych&Vaneeva&Ivanova2007,Vaneeva&Popovych&Sophocleous2009},  
B\"acklund transformations etc. 
For example, given a ``spectral problem''~$\mathcal P$: $\psi_x=\mathcal U\psi$, $\psi_t=\mathcal V\psi$ associated with an equation~$\mathcal L$ 
(with no boundary conditions), a point transformation~$\mathcal T$: $\tilde t=T(t)$, $\tilde x=X(t,x)$, $\tilde u=U(t,x,u)$ 
with $T_tX_xU_u\ne0$ maps~$\mathcal P$ to the system~$\tilde{\mathcal P}$: 
$\psi_{\tilde x}=\tilde{\mathcal  U}\psi$, $\psi_{\tilde t}=\tilde{\mathcal V}\psi$, 
where $\tilde{\mathcal U}=X_x{}^{-1}\mathcal U$ and $\tilde{\mathcal V}=T_t{}^{-1}(\mathcal V-X_x{}^{-1}X_t\mathcal U)$, 
whose compatibility condition~$\tilde{\mathcal L}$ is similar to~$\mathcal L$ with respect to~$\mathcal T$. 
At the same time, even similar equations can demand a separate investigation, e.g., of related boundary problems and asymptotic properties
\cite{Matveev&Salle1991}.

The next paper of the series will be devoted to the ninth common error.
We plan to exemplify the discussion on finding exact solutions of linearizable differential equations 
with papers on equations from the Burgers hierarchy, their different generalizations and multidimensional versions.

\subsection*{Acknowledgements}

The authors are grateful to N.~Ivanova, A.~Sakhnovich and C.~Sophocleous
for helpful remarks and relevant references.
The research of ROP was supported by project P20632 of the Austrian Science Fund.


\end{document}